\documentclass[reprint,twocolumn,aps,prc,a4paper,superscriptaddress,showpacs,preprintnumbers,amsmath,amssymb]{revtex4}
\usepackage{graphicx}
\usepackage{dcolumn}
\usepackage{bm}
\usepackage{CJK}
\usepackage{float}
\usepackage{subfigure}

\begin{document}

\title{New charge radius relations for atomic nuclei}

\author{B.H. Sun}\thanks{Corresponding author: bhsun@buaa.edu.cn}
\affiliation{School of Physics and Nuclear Energy Engineering, Beihang University, Beijing 100191, China}
 \affiliation{International Research Center for Nuclei and Particles in the Cosmos, Beijing 100191, China}
\author{Y. Lu}
\affiliation{Department of Physics and Astronomy, and Shanghai Key Lab for Particle Physics and Cosmology, Shanghai Jiao Tong University, Shanghai, 200240, China}%
\author{J.P. Peng}
\affiliation{School of Physics and Nuclear Energy Engineering, Beihang University, Beijing 100191, China}
\author{C.Y. Liu}
\affiliation{School of Physics and Nuclear Energy Engineering, Beihang University, Beijing 100191, China}
\author{Y.M. Zhao}\thanks{ymzhao@sjtu.edu.cn}
 \affiliation{Department of Physics and Astronomy, and Shanghai Key Lab for Particle Physics and Cosmology, Shanghai Jiao Tong University, Shanghai, 200240, China}%

\date{\today}
\begin{abstract}
We show that the charge radii of neighboring atomic nuclei, independent of atomic number and charge, follow remarkably very simple relations, despite the fact that atomic nuclei are complex finite many-body systems governed by the laws of quantum mechanics. These relations can be understood within the picture of independent-particle motion and by assuming that neighboring nuclei have similar patterns in the charge density distribution.
A root-mean-square (rms) deviation of 0.0078 fm is obtained between the predictions in these relations and the experimental values, i.e., a precision comparable  modern experimental techniques.
Such high accuracy relations are very useful to check the consistence of the nuclear charge radius surface and moreover to predict unknown nuclear charge radii, while large deviations from experimental data are seen to reveal the appearance of nuclear shape transition or coexsitence.

\end{abstract}
\maketitle
\date{today}

Like many systems governed by the laws of quantum mechanics, the nucleus is
an object full of mysteries whose properties are much more difficult to characterize than those of macroscopic objects.
Rather than build an exact replica of the nuclear system (almost an impossible job), nuclear physicists in reality
have selected a different approach, using a relatively small number of measurable properties of quantum systems
to specify the overall characteristic of the entire nucleus. Investigations of these properties and moreover using them as bridges to reveal the atomic nucleus, form the basis of present nuclear physics investigations.
Nuclear extension in space, often characterized by charge radius, is one of such static properties.

Nowadays, nuclear size data~\cite{Angeli13} constitute one of the most precise and extensive arrays of experimental
information available for the interpretation of nuclear phenomena.
The pioneering works by using various methods such as electron scattering and muonic atomic spectroscopy~\cite{Fricke95},
indicated that a nuclide near the $\beta$-stability line behaves like a solid sphere with constant density.
In recent decades, the enormous progress in size determination of exotic nuclides has been entwined with the realization of radioactive ion beams~\cite{Ozawa01} and fast developments especially in ultra-high-sensitivity laser spectroscopy techniques ~\cite{Cheal10,Blaum13}.
The new measurements contributed to revealing, for example, the neutron skin effect~\cite{Yamaguchi11,Ozawa14} and the nuclear shape variances
~\cite{Seliverstov13,Flanagan13,Budin14} when moving far away from the stability line.
These new results are among the strongest motivations for the next generations of nuclear physics facilities, in which
electron-scattering experiments~\cite{Suda09,ELISE11} on unstable nuclei are under way.
In the march towards the new era of nuclear physics, the knowledge of nuclear sizes plays a very important role in understanding complex atomic nuclei.

New precision data in turn challenge the present theories and trigger innovations to interpret the results and to understand underlying nuclear structures.
A conventional way to estimate the nuclear charge radii was to improve the empirical formula of the $A^{1/3}$ law, e.g., by including residual corrections lsuch as isospin and shell corrections~\cite{Wang13}, or by introducing the $Z^{1/3}$ dependence~\cite{Zhang02}. Here $A, Z$ denote the nuclear mass and proton number, respectively.
In nuclear density functional theories  based on the
mean-field approach such as the Hartree-Fock-Bogoliubov (HFB) model~\cite{Goriely09,Goriely10} and the relativistic mean-field (RMF) theory~\cite{Geng03,Zhao10}, nuclear charge radii are calculated in a self-consistent way by folding the charge density distribution.
Besides,  recent work attempts to deduce charge radii based on the $\alpha$ decay~\cite{Ni13,Qian14}, and cluster and proton emission data~\cite{Qian13}.
In this work, however, different from the above approaches, we
propose a set of new difference equations of charge radius for neighboring nuclei, independent of atomic number and charge. Such simple relations are expected to have good accuracies due to resultant cancellations of the proton distributions (and also neutrons of relevance) in space.

The first question considered is whether the differences
between nuclear charge radii can be understood based on a smooth evolution of nuclear size, or on the basis of well-known knowledge about the nuclear many-body system. Similar to nuclear mass relations such as the Garvey-Kelson mass relations~\cite{GK69}, a necessary condition for such cancellations is that the number of neutrons and protons and of the various pairs cancels in this relations.

With one proton added, the charge radius increases; with one more neutron added, the $np$ interaction also leads to a change of the charge radius. In the framework of the independent particle shell-model, there holds the following relation:
\begin{eqnarray}
   \delta R_{1p-1n}(Z,N)&= &R(Z,N)+R(Z-1,N-1)     \nonumber \\
                        & &-R(Z,N-1)-R(Z-1,N)    \nonumber \\
                        &\simeq & 0  \;,
   \label{eq1}
\end{eqnarray}
where $R(Z,N)$ is the charge radius of a nucleus with $N$ neutrons and $Z$ protons. The nuclear charge radius data are taken from Ref.~\cite{Angeli13}.

\begin{figure}
\centering
\includegraphics[angle=0, width=0.4\textwidth]{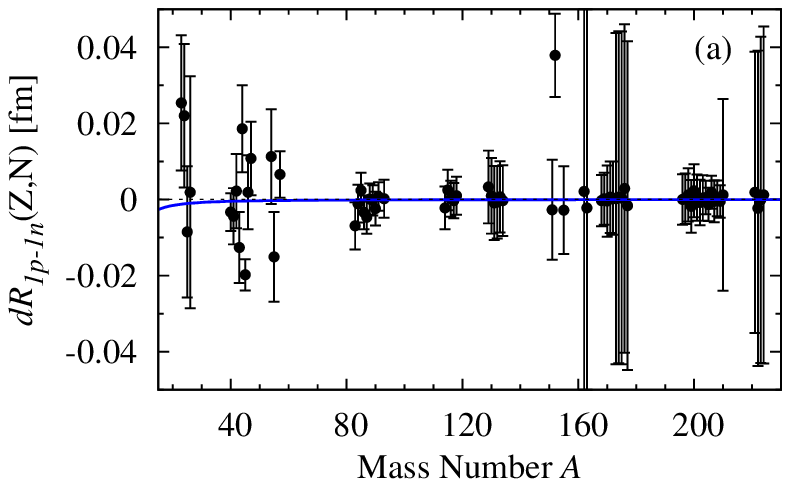}
\includegraphics[angle=0, width=0.4\textwidth]{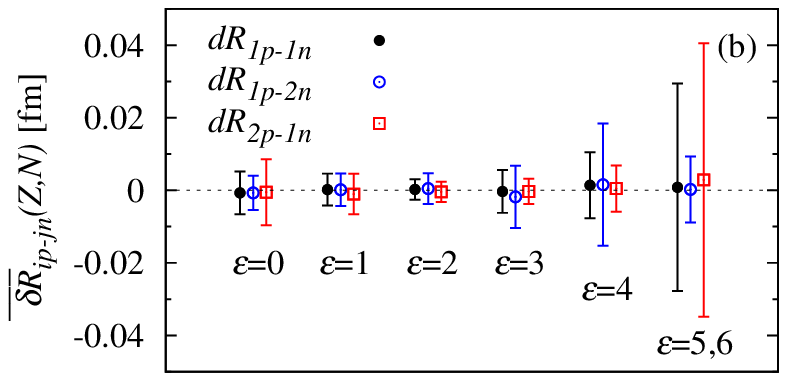}
\caption{(Color online) (a) $\delta R_{1p-1n}$(Z,N) for most stable nuclides.
The blue solid line represents the $\delta R_{1p-1n}$(Z,N) deduced with the empirical $A^{1/3}$ formula.
(b) Dependence of the mean value of $\delta R_{ip-jn}$(Z,N), $\overline{\delta R}_{ip-jn}$(Z,N), together with its uncertainty, on
the isobaric distance from stability $\varepsilon$.}
\label{fig1}
\end{figure}

The validity of this relationship is readily discernible in Fig.1(a) for most stable nuclides.
It is found that the deviations from zero are indeed small and essentially random. The weighted mean values
$\overline{\delta R}_{1p-1n}$ amount to only -0.7$\times 10^{-3}$ fm with the weighted root-mean-square (rms) deviation of
 5$\times 10^{-3}$ fm. The uncertainties of $\delta R_{1p-1n}(Z,N)$, arising from the experimental uncertainties of four nuclides involved, are typically about $10^{-2}$ fm, and moreover the cases with the largest uncertainties are located at either the isotopic chains Dy and Ho around $N=90$, or Ir and Pt around $Z=82$.

Especially, we would like to draw attention to the cases with magic numbers $Z=50/82$ and/or $N=50/126$, whose relevant $\delta R_{1p-1n}$ of less than 2.5$\times 10^{-3}$ fm are remarkably consistent with others. This striking feature is also elucidated in Fig.~\ref{fig4}.
The only exception for intermediate-heavy nuclides  that falls out of the one standard deviation is $\delta R_{1p-1n}(63,89)$. This is due to the
abrupt changes in nuclear shape around $N \sim 90$, and we will come back to this point later.
Comparing with the heavy nuclei, a relatively large scatter around zero is shown for the light mass systems with $A<60$. Nevertheless, this is as expected from the nature of the ``few-body'' complexity, while the hypothesis of slow variance of nuclear charge radius with increasing nucleons  should hold better for the ``mean-field" picture of heavy systems.

For illustrative purpose, Fig.1(a) displays also the results computed based on the empirical formula $R=0.956 A^{1/3}$, which is obtained by fitting to the rms charge radii of 296 stable isotopes for 83 elements.
With increasing mass number, $\delta R_{1p-1n}(Z,N)$ deduced from this $A^{1/3}$ formula
goes down to a few times $10^{-5}$ fm.
Another empirical formula, the $Z^{1/3}$ dependence, would lead to exactly zero.
This fact demonstrates that Eq.~(\ref{eq1}) is fulfilled nicely in the
consideration of the nuclear saturation property, one of the main ingredients for modern nuclear theories.

To examine how well $\delta R_{1p-1n}(Z,N) \approx 0$ can reach in precision when going far away from stability, for each isobaric chain with mass $A$, we parametrize the distance between the nuclide $(Z,A-Z)$ and the nuclide $(Z_0,A-Z_0)$ in the $\beta$-stability line~\cite{Sun08} by the isobaric distance from stability $\varepsilon=|Z_0-Z|$ with
  \begin{eqnarray}
   \label{eq:stable}
   Z_0=\frac{A}{1.98+0.0155A^{2/3}} \; .
  \end{eqnarray}
Thus, $\varepsilon=0$ stands for the most stable nuclei.
Figure~\ref{fig1}(b) shows the behavior of $\delta R_{1p-1n}(Z,N)$ as a function of $\varepsilon$.
The mean deviations $\overline{\delta R}_{1p-1n}$(Z,N) are very small and in the same order of magnitude for larger $\varepsilon$.
The displayed uncertainties of typically 5$\times 10^{-3}$ fm, include both the experimental uncertainties
from Eq.~(\ref{eq1}) and the weighted rms deviations
from the mean values, while the latter are ten times larger than the experimental uncertainties.
Only at $\varepsilon \geqslant 4$, i.e., for the most exotic cases in the known charge radius surface,
is there a slightly large deviation from the mean value.  This deviation is partially due to the large experimental uncertainties, about one order of magnitude more than those close to stability line.

The success of this formula lies in the fact that the density of nucleons in the inner regions of all nuclei (especially neighboring nuclei) is about the same and that the surface thicknesses of all nuclei are very similar and vary slowly in the nuclear chart.
Indeed, Eq. (\ref{eq1}) represents the mixed partial derivative of nuclear charge radius surface with respect to $N$ and $Z$, i.e.,
\begin{equation}
  \delta R_{1p-1n}(Z,N) \approx \frac{\partial R^2(Z,N)}{\partial N\partial Z} \;,
  \label{eq3}
\end{equation}
while the first-order differences, $\partial R(Z,N)/\partial N$ and $\partial R(Z,N)/\partial Z$ are canceled in this relation. To see  to what extent the above equation works over the nuclear chart, one may resort to the
 the empirical formula $R_{ch} \propto A^{1/3}$. The $\delta R_{1p-1n}(Z,N)$ in this special case is
proportional to $A^{-5/3}$, thus it amounts to only a few times of $10^{-4}$ fm when $A$ equals to 100.
The relatively larger deviations from zero at lighter nuclei can also be understood
because of $A^{-5/3}$ factor.  

Recursive application of Eq. ~(\ref{eq1}) yields a general formula,  $\delta R_{ip-jn} \simeq 0$, where $i$ and $j$ are integers. Two simple cases are as follows.
\begin{eqnarray}
   \delta R_{1p-2n}(Z,N)&= &R(Z,N)+R(Z-1,N-2)     \nonumber \\
                        & &-R(Z,N-2)-R(Z-1,N)    \nonumber \\
                        &\simeq & 0  \;, \nonumber \\
   \delta R_{2p-1n}(Z,N)&= & R(Z,N)+R(Z-2,N-1) \nonumber \\
                        & & -R(Z,N-1)-R(Z-2,N) \nonumber \\
                        &\simeq & 0  \;.
   \label{eq5}
\end{eqnarray}

The results of these two relations are also summarized in Fig. 1(b).

Equations (\ref{eq1})  and (\ref{eq5}) offer a very useful feature in prediction of the rms charge radius of a given nucleus in terms of charge radii of its three neighboring nuclei.
This is an alternative way to examine the high precision of the three relations. Taking Eq.~(1) as an example, this calculation can be done in four different forms, as we choose any of the four to be evaluated from the others. Using all these three formulas, one can then have a maximum of 12 predictions for the rms charge radius of a given nucleus, provided all neighboring charge radius are known; otherwise, the number of predictions $n$ is $n < 12$. If predictions based on mass relations are essentially random with respect to zero, a predicted result
obtained by averaging all available predictions is expected to be more accurate, as suggested and discussed in Refs. ~\cite{Barea05,Barea08}.
Similar features are also seen in the results of charge radii here. As tabulated in Table I, there are only 17 nuclei with known charge radii that can be predicted in 12 different ways, while more than 600 nuclei can be predicted in more than two different ways.  The rms deviations are around 0.004 fm for $n \geqslant 5$. Note that the farther a nuclide is from the $\beta$-stability line, the fewer  possible predictions ($n$) one has for this nuclide.

\begin{center}
\begin{table}[htp]
\centering
\caption{Weighted rms ($\sigma$) and mean ($\bar{\epsilon}$) deviations (in $10^{-3}$fm) between experimental data and predictions in this work, when different number of predictions $n$ exist for a given nucleus. $N$ refers to the total number of nuclei that can have $n$ possible predictions. Only nuclei with $N \geqslant 2$ and $Z \geqslant 2$ are considered.}
\begin{tabular}{|c|rrr|r|rrr|}
\hline
$n$	& $N$ &  $\bar{\epsilon}$   & $\sigma$ & $n$	& $N$ &  $\bar{\epsilon}$   & $\sigma$ \\
\hline
1		&  76   &  0.22    &   13.05     & 7	&  58   & 0.49    &   7.35   \\
2		&  88   & -0.28   &   14.72     & 8	&  81   & 1.00    &   4.30   \\
3		&  66   &  0.19    &   6.42      & 9	&  34   & -0.54    &   4.16  \\
4		&  88   & -1.65   &   13.33     &10	&  53   & -0.37    &   4.33   \\
5		&  55   &  0.16    &   3.95      &11	&  19   & -0.96    &   4.19   \\
6		&  152  & 0.29    &   4.12      &12	&  17   & 1.18    &   3.48   \\
\hline		
\end{tabular}
\label{tab2}
\end{table}
\end{center}

\begin{figure}
\centering
\includegraphics[width=0.45\textwidth, angle=0]{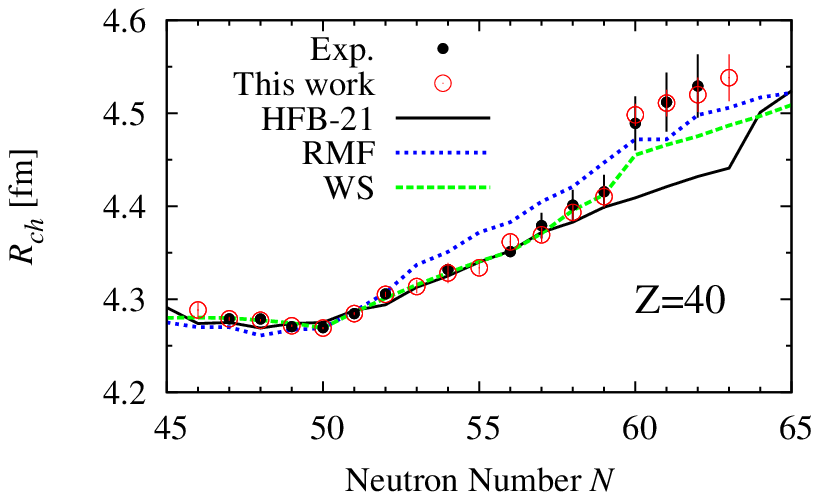}
\includegraphics[width=0.45\textwidth, angle=0]{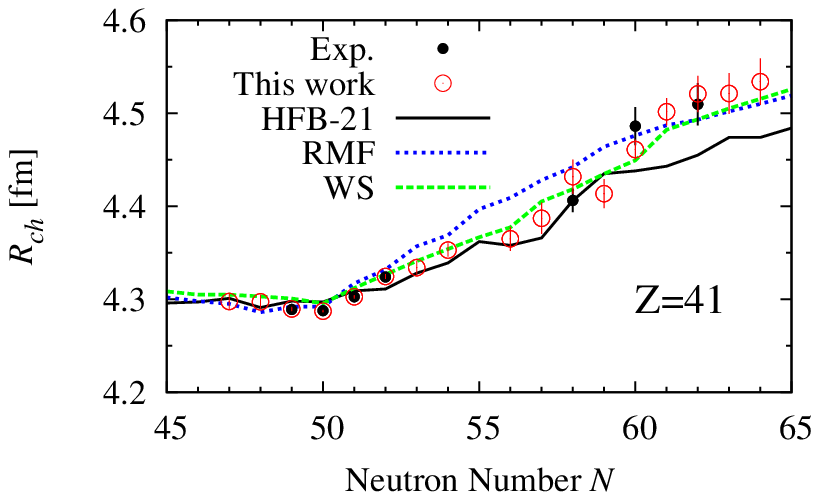}
 \caption{Comparisons of predictions in nuclear charge radius and available experimental data for Z=40 (a) and 41 (b) isotopic chains.}
\label{fig2}
\end{figure}

 \begin{figure*}
\centering
\includegraphics[angle=0, width=0.7\textwidth]{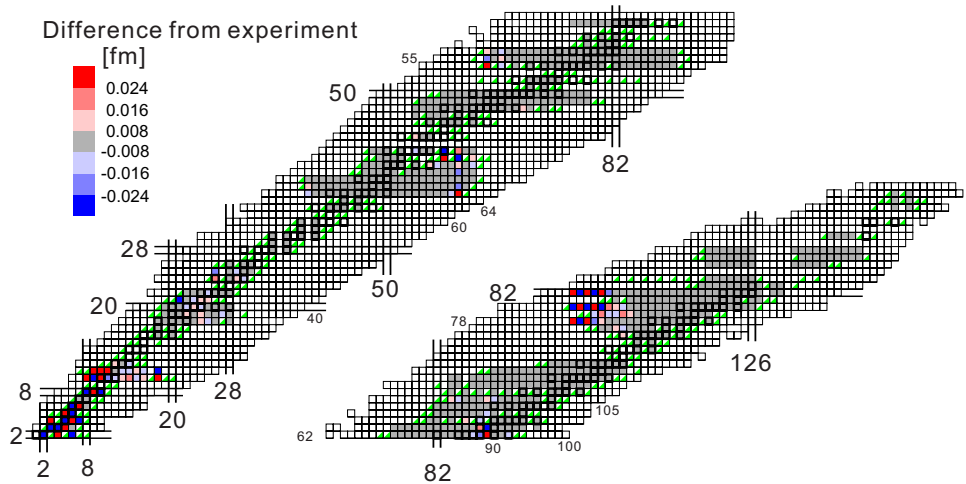}
 \caption{(Color online) Chart of the nuclides showing the differences (color coded, in fm) between the predictions and experimental values.
The black squares indicate stable isotopes. The magic numbers are displayed by pairs of parallel lines.
325 nuclides whose rms charge radius can be predicted within an accuracy around 0.01 fm are marked  by the green triangles.
 }
\label{fig3}
\end{figure*}

When several predictions ($n>1$) exist for a specific nuclide, the mean value of different predictions is then calculated and treated afterwards as the prediction for this nuclide. Eventually, one can obtain a set of 788 ``new" rms charge radius values, and accordingly their differences from experimental data. The deviations from experiments are summarized in Table~\ref{tab2}.
The results again demonstrate the reliability of Eqs.~(\ref{eq1}) and (\ref{eq5}). The larger rms values for $n \leqslant 4$ are mainly caused by the nuclei with $Z<20$ and $Z\approx 82$.
Globally the rms deviation on charge radii amounts to 0.0078 fm for all the nuclei with $N, Z\geqslant 2 $ in this work, i.e.,
an accuracy comparable to the experimental data and, to our knowledge, by far better than
any global nuclear models and empirical formula available.
A rms value of more than 0.02 fm is obtained for
the up-to-date microscopic-rooted nuclear models~\cite{Goriely09,Goriely10,Geng03,Zhao10} and a recent proposed four-parameter formula including the shell corrections and deformations of nuclei obtained from the Weizs\"{a}cker-Skyrme (WS) mass model~\cite{Wang13}. When considering only charge radii of nuclei with $Z\geqslant 20$, the global rms value deceases to 0.0069 fm, while it increases to 0.0194 fm for light nuclei with  $Z< 20$.

More insight into the high precision of the relations predictions is shown in Fig.2, in which the available experimental data and our predictions are plotted for both Zr and Nb isotopes.  For the sake of comparison, three up-to-date model predictions, RMF~\cite{Geng03}, HFB-21~\cite{Goriely10}, and WS~\cite{Wang13} are also included.  The consistency of our predicted and experimental trends is clearly verified, even in the shape transition region around $N\approx 60$. In cases of experimental charge radii being unavailable, the predictions follow the smooth evolution along the isotopic chain. These smooth evolutions along isotopic chains have also been seen in other nuclear observables, e.g., mass~\cite{Audi2012}, when no sudden changes occur in neighboring atomic nuclei. The HFB-21  nicely reproduces the nuclear charge radii before the shape transition at $N \approx 60$, but cannot give the trend for $N>60$, while the RMF tends to show overestimated values for nuclei with neutron number large than the magic number 50. The WS model, on the other hand, can generally reproduces very well the kinks at $N=60$ in both the isotopic chains.  

The global comparisons of the relations predictions with experimental data are shown in Fig.~\ref{fig3} .
Several features of this chart are worthy of note. (1) Equations~(\ref{eq1}) and (\ref{eq5}) hold remarkably well throughout  the nuclear chart, especially concerning the nuclides heavier than Ni.
(2) For light nuclides from He to Ne, the deviations can be as large as 0.1 fm. As already mentioned, this is due to the few body nature of these systems. The strong odd-even staggering effect indicates that some important effects fail in a large extent to be washed out. Dedicated
examinations on the charge density distribution of light nuclides would be very helpful to understand this phenomenon.
(3) No conclusive evidence yet exists indicating the possible failure of Eqs.~(\ref{eq1}) and (\ref{eq5})  at neutron magic numbers 50/82/126 and proton numbers 50/82,  for available experimental data. In other words, available data tend to support the reliability of Eqs.~(\ref{eq1}) and (\ref{eq5}) around those shells.  This is quite different from the local nuclear mass relations, like the valence proton-neutron interactions~\cite{Cakirli05,Stoitsov07,Chen10}, where these empirical interactions are sensitive to changes in shell structure in exotic nuclei. (4) The significant local variations
are seen in the intermediate and heavy elements at $N \sim 60$, $N \sim 90$ and $Z \sim 80$. The slight increased cases also include the neutron-deficient nuclei at $Z \sim 55$ and $N\sim 40$.
These groups are normally quoted as typical ``shape phase transition" or ``shape coexistence" zones~\cite{Angeli09,Cejnar10,Heyde11}. Sudden changes in the shape and single-particle distributions with the number of protons and neutrons result in
a large residuals other than zero.
Investigations on the relevant experimental data combining with other observables such as mass and spectroscopic (see, e.g. Ref.~\cite{Cakirli10}), clearly show that the large deviations from experiments are strongly correlated with sudden nuclear shape variances.
In Ref.~\cite{Angeli09} the ``kink strength", the second difference of charge radius versus nucleon number $d^2R(Z, N)/dN^2$ or $d^2R(Z, N)/dZ^2$, is defined to characterize the changes in nuclear structure due to shell effects and deformation.
Here we show that $\delta R_{pn}$ can be used as a good probe to characterize the changes in nuclear charge radius surface caused by shape transition or shape coexistence.

\begin{figure}
\centering
\includegraphics[scale=0.65, angle=0]{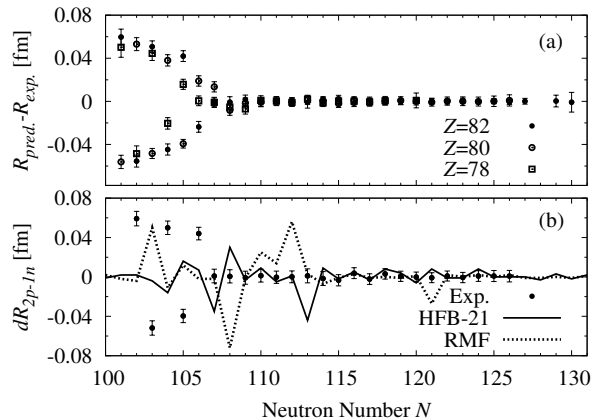}
 \caption{(a) Differences between our predictions and experimental data for isotopic chains of Pb, Hg and Pt.
 (b) $\delta R_{2p-1n}$ in the RMF and HFB-21 mass models with experimental data for Pb isotopes.}
\label{fig4}
\end{figure}

Figure~\ref{fig4}(a) explicitly displays our predictions at $Z \sim 82$. One sees an increasing deviation from zero for $N<108$. This is consistent with the onset of shape coexistence of deformation at neutron number around 108~\cite{Witte07}. A closer scrutiny of our predictions shows that the differences are mainly determined by $\delta R_{2p-1n}=0$. The anomalous variances for $N<108$ is due to the huge odd-even staggering observed in the light mercury region~\cite{Ulm86}. Accordingly, we plot in Fig.~\ref{fig4}(b) the results of $\delta R_{2p-1n}$ for Pb isotopes. Illustrated in this figure are also the calculated $\delta R_{2p-1n}$ in the HFB-21~\cite{Goriely10} and RMF models~\cite{Geng03}. Both models cannot reproduce the experimental staggering in the light Pb isotopes in both magnitude and sign of the odd-even staggering.

In summary, we propose a set of nuclear charge radius relations assuming that the neighboring nuclei
share the similar pattern in the charge density distribution. This assumption is the main ingredients for the success of the nuclear shell model or generally the mean field approaches.
In this respect, the new relations work better for heavier nuclear system as shown in Fig.~\ref{fig3}. We have shown that the precision of the proposed new relations holds for not only stable nuclei but also the most exotic cases, as demonstrated in Fig.~\ref{fig1}.
The significance of this work is that it offers us  {\it a simple} but {\it very accurate} way to investigate the consistence of the charge radius surface (in both experimental data and predictions of nuclear models). This is particularly important
considering the fact that there are many sources of charge radius data deduced from very different experimental techniques. After all, the difference in radii of neighboring isotopes is very small, from about 0.1 fm for $A \sim 10$ down to about 0.02 fm for $A \sim 100$. 
Moreover, the proposed formula can be naturally used to predict unknown nuclear charge radii. For instance, the charge radii of 325 nuclides can be estimated as indicated in Fig.~\ref{fig3} within an accuracy of typically 0.01 fm.

The available data tend to support the reliability of our proposed relations even at neutron magic numbers 50/82/126 and proton numbers 50/82.  This success might be due to the interplay of neutrons and protons in our formula, which
somehow washes out the shell effect. Accordingly, it is very interesting to see how well the new relations hold towards drip lines, where one expects the disappearance of the traditional magic numbers and/or the appearance of new magic ones. 
Any new experimental data towards the drip lines are thus very valuable to verify the predictive power of the relations.

The large deviations at, e.g., the $Z \sim 82, N \sim 100$ region, 
suggesting sudden, very large variations in charge density distributions for relevant nuclei,
are pointed out to be well correlated with the appearance of shape transition or coexistence. Therefore this provides us with a {\it new} and {\it reliable probe} for sudden changes in nuclear shape.
Accordingly, when sudden variances occur in nuclear shape or are expected, one has to be cautious about the applications of Eqs.~\ref{eq1} and \ref{eq5} to unknown charge radii.
In these cases, it may be instructive to check the correlations with other ground and excited state observables.

Finally we wish to note without details a recent work in which the Garvey-Kelson mass relations were extended to nuclear charge
radii~\cite{Pieka10}.  Despite fewer nuclei involved, our relations show similar precisions in reproducing known nuclear charge radii.

\acknowledgments

This work was supported partially by the National Natural Science Foundation of China
(No. 11035007, 11105010, 11128510, 11235002, 11225524 and 11475014), the 973 Program of China (Grant No. 2013CB834401), Shanghai Key Laboratory (Grant No. 11DZ2260700) and the Fundamental Research Funds for the Central Universities. We would like to thank I. Angeli, and K. Marinova for providing
the rms charge radius data, K. Blaum, P.W. Zhao, and S.G. Zhou for stimulating discussions, and the anonymous referee  for useful comments.


\end{document}